\documentclass[twocolumn,smallextended]{svjour3}
\smartqed  
\usepackage{graphicx,dcolumn,bm,color,latexsym,amssymb,amsmath}
\usepackage[latin1]{inputenc}
\usepackage[english]{babel}
\usepackage{wrapfig,times}
\usepackage{amsfonts}
\usepackage{verbatim}

\definecolor{Blue}{rgb}{0.0,0.0,1}
\definecolor{Red}{rgb}{1,0.0,0.0}

\begin{document}

\title{Experimental implementation of a NMR entanglement witness}

\author{J. G. Filgueiras         \and
        T. O. Maciel         \and
        R. E. Auccaise         \and
        R. O. Vianna         \and
		R. S. Sarthour         \and
		I. S. Oliveira
}

\institute{J. G. Filgueiras \and R. S. Sarthour \and I. S. Oliveira\at
              Centro Brasileiro de Pesquisas F\'{\i}sicas, Rua Dr.
Xavier Sigaud 150, Rio de Janeiro 22290-180, RJ, Brazil\\
              \email{jgfilg@cbpf.br}           
           \and
           T. O. Maciel \and R. O. Vianna \at
				Departamento de F\'{\i}sica - ICEx - Universidade Federal de Minas Gerais,
Av. Pres.  Ant\^onio Carlos 6627, Belo Horizonte 31270-901, MG - Brazil
		   \and
		   R. E. Auccaise \at
                Empresa Brasileira de Pesquisa Agropecu\'aria, Rua Jardim Bot\^anico 1024,
Rio de Janeiro 22460-000, RJ - Brazil
}			

\maketitle

\begin{abstract}
Entanglement witnesses (EW) allow the detection of entanglement in a quantum system, from the measurement of 
some few observables. They do not require the complete determination of the 
quantum state, which is regarded as a main advantage. On this paper it is experimentally analyzed 
an entanglement witness recently proposed in the context of Nuclear Magnetic Resonance (NMR) experiments to test it in some Bell-diagonal states.
We also propose some optimal entanglement witness for Bell-diagonal states. The efficiency of the two types of EW's are compared 
to a measure of entanglement with tomographic cost, the generalized 
robustness of entanglement. It is used a GRAPE algorithm to produce an entangled state which is out of the detection region 
of the EW for Bell-diagonal states. Upon relaxation, the results show that there is a region in which both EW fails,
whereas the generalized robustness still shows entanglement, but with the entanglement witness proposed here with a better performance.
\end{abstract}
\section{Introduction}

Entanglement is one of the central keys for quantum information processing, being associated with various 
puzzling quantum phenomena, such as Bell's inequality violation and quantum teleportation, for example. It 
is also a main resource for the exponential speedup in some quantum algorithms \cite{MNielsen}. Therefore,
the detection of entanglement is important. For that, many tools have been developed, one of these are the
entanglement witnesses.

Entanglement witnesses are tools designed to detect entanglement from direct measurements of 
observables. They can be used either to detect entanglement in a given state, or to
quantify the entanglement for a specific state or class of states \cite{Lewenstein}.
The main advantage of the use of entanglement witnesses is the possibility of detection of entanglement
without performing Quantum State Tomography, which can significantly reduce the number of measurements performed in the system, 
in order to characterize some quantum effect due to the presence of entanglement.

During the last few years there have been proposals and experimental implementations of entanglement witnesses for 
various different quantum systems, like optical \cite{Saavedra} and magnetic ones 
\cite{Vedral, Rappoporst, Alexandre}. Although NMR techniques have been successfully used 
to implement quantum protocols \cite{Oliveira}, like quantum teleportation \cite{Nielsen} and Shor's \cite{Chuang} algorithm, 
besides several other quantum simulations \cite{Oliveira}, only recently a proposal of EW has been made for 
NMR quantum information processing experiments \cite{Rahimi}. In this paper it is demonstrated 
the implementation of such a EW in a class of states, and it is also proposed some other entanglement witnesses.

\section{Theory}

\begin{center}
\begin{figure}%
\includegraphics[scale=0.6]{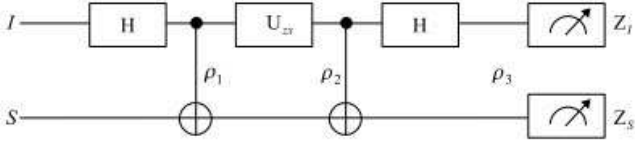}
\caption{Quantum circuit for superdense coding. In the first part of the circuit, the input state $|00\rangle$ being transformed 
in the cat state, $|\Phi^+\rangle$. After it, the ``message'' operator is applied, taking $|\Phi^+\rangle$ 
into one of the four Bell basis states. The final part of the circuit indicates the measurement in the Bell basis.}%
\label{superdense}%
\end{figure}
\end{center}

Recently, a proposal for an EW has been made for NMR quantum information 
processing \cite{Rahimi} which uses the superdense coding protocol, successfully implemented 
with NMR by Fang et al \cite{Fang}. The circuit for the superdense coding is given in the Figure \ref{superdense}.

For a pure state, the circuit transmits two classical bits of information with only one qubit transmitted.
The input state in the superdense coding, $|00\rangle$, passes through an ERP gate \cite{Oliveira}, becoming the cat state, 
$|\Phi^+\rangle$ (first part of Figure \ref{superdense}). The encoded message is then chosen by a ``message'' 
operator, applied only at the first qubit, that transforms the cat state into one of the four states of the Bell 
basis (the operator $U_{xz}$ at the second part of Figure \ref{superdense}). This operator is given by $X$, $Z$ or the product $XZ$.
Then, the first qubit, which was modified by the message operator is sent to the other person which 
has the other qubit of the entangled pair, and a measurement at the Bell basis is performed in each qubit (final part of the 
Figure \ref{superdense}). The result of the readouts, measured at each qubit, is dependent of the ``message'' operator. 
By the knowledge of the sent message, the transmission of the two classical bits of information with only one qubit 
transmited is completed.

In NMR systems one deals with not pure, but mixed states, due to the large number of molecules in a sample.
Then, it is necessary to consider the above circuit in the context of mixed states. 
The equilibrium state of a NMR system containing two qubits can be written in form
\begin{equation}\label{01}
\rho = (p_I|0\rangle\langle 0| + q_I|1\rangle\langle 1|)\otimes (p_S|0\rangle\langle 0| + q_S|1\rangle\langle 1|),
\end{equation}
\par\noindent
where 
\begin{equation}\label{02}
p_i = \frac{1+\epsilon_i}{2}, \qquad q_i = \frac{1-\epsilon_i}{2},
\end{equation}
\par\noindent
and the indexes $I$ and $S$ label each of the nuclear spins used as qubits. 

By applying an EPR gate \cite{Oliveira} (first part of the circuit in Figure \ref{superdense}) to this state
the output will be a Bell-diagonal state:
\begin{eqnarray}\label{03}
\rho_1 &=& p_{I}p_{S}|\Phi^+\rangle\langle\Phi^+| + p_{I}q_{S}|\Psi^+\rangle\langle\Psi^+| \cr
&+& p_{S}q_{I}|\Phi^-\rangle\langle\Phi^-| + p_{I}p_{S}|\Psi^-\rangle\langle\Psi^-|,
\end{eqnarray}
\par\noindent
where $|\Phi^+\rangle$, $|\Psi^+\rangle$, $|\Phi^-\rangle$ and $|\Psi^-\rangle$ are the states of the Bell basis. The parameter $\epsilon_i$ 
is the relation between the magnetic and thermal energies and is typically $\approx 10^{-5}$.

The spin magnetizations are measured at the end of the circuit and are proportional to the polarization of the state, $\epsilon$ (see the section 4). They are given by
\begin{equation}\label{04} 
\left\langle Z_I\right\rangle = (-1)^{z}\epsilon_I, \qquad \left\langle Z_S\right\rangle = (-1)^{x}\epsilon_S.
\end{equation}
\par\noindent
If the variables $x$ and $z$, the encripted message, are known, the NMR implementation of the superdense coding is successful. 

The statistical condition for success in the NMR implementation of superdense coding and the condition for the entanglement of the Bell-diagonal states 
are the same, and are given by \cite{Rahimi}:
\begin{equation}\label{05}
p_{I}p_{S} > \frac{1}{2}
\end{equation}
This equation can be used to define an entanglement witness, as $F$ $=$ $1/2 -p_{I}p_{S}$. Using the expressions for the 
probabilities and those for $p_i$ and $q_i$, we have
\begin{equation}\label{06}
F = \frac{1}{2} - \frac{1}{4}(1+|\left\langle Z_I\right\rangle|)(1+|\left\langle Z_S\right\rangle|).
\end{equation}
The measurements of the magnetizations $Z_I$ and $Z_S$ at the end of the circuit are equivalent to the measurements 
of $\rho_1$ (see Fig. \ref{superdense}) in the Bell basis, since
\begin{equation}\label{07}
\left\langle Z_I\right\rangle = Tr(\rho_{f}(Z_{I}\otimes \mathbb{I}_S)
= Tr(\rho_{1}(X_{I}\otimes X_{S}) \equiv \left\langle W_1\right\rangle,
\end{equation}
\begin{equation}
\left\langle Z_S\right\rangle = Tr(\rho_{f}(\mathbb{I}_I\otimes Z_S)
= Tr(\rho_{1}(Z_{I}\otimes Z_{S}) \equiv \left\langle W_2\right\rangle,
\end{equation}
which yields:
\begin{equation}\label{08}
F = \frac{1}{2} - \frac{1}{4}(1+|\left\langle W_1\right\rangle|)(1+|\left\langle W_2\right\rangle|).
\end{equation}
\par\noindent This equation shows explicitly that $F$ is a measure of the correlations between the two qubits.

\section{Decomposable optimal entanglement witnesses for NMR}

In this section, we propose a set of optimal decomposable entanglement 
witnesses, which can detect the entanglement of  states in the 
vicinity of the Bell states. In relation to the witness $F$, one
pays the price of performing just one more local measurements, in
order to have a finer description of the entanglement, as seen in
Fig. 7 (see next section).

The new witnesses  are of the form:
\begin{equation}
\label{wxyz}
W=C_I\mathbb{I}+C_x X_I\otimes X_S + C_y Y_I\otimes Y_S + C_z Z_I\otimes Z_S.
\end{equation} 
To guarantee that $W$ have a positive expectation value on all
separable states $\sigma$, we have just to impose that the
partial transpose of $W$ is a positive operator, i.e.,
$W^{T_A}\geq 0$. This follows from the fact that if $\sigma$ is
a bipartite separable state, so is its partial transpose, $\sigma^{T_A}$.
Therefore $Tr(W \sigma^{T_A})=Tr(W^{T_A} \sigma)\geq 0$, which shows that
$W$ is a valid entanglement witness.

Now, for a given Bell state $|\beta_{ii}\rangle$, an optimal witness in the form of Eq.(\ref{wxyz}) 
is obtained by solving  the following semidefinite program (sdp):
\cite{sdp,Brandao}:
\begin{center}
{\em minimize} $\langle\beta_{ii}| W |\beta_{ii}\rangle$
\begin{equation}
\label{sdpw}
subject\,\, to 
\left\{
\begin{array}{l}
 W^{T_A}\geq 0, \\
W\leq \mathbb{I}.
\end{array}
\right.
\end{equation}
\end{center}
The sdp (Eq.\ref{sdpw}) yields the witnesses in Tab. 1.
\begin{center}
\begin{table}\label{tabela}
\begin{tabular}{l|c|c|c|c} \hline
 & I & XX & YY & ZZ \\ \hline
$|\Phi^+\rangle$ & 0.5 & -0.5 & 0.5 & -0.5 \\ \hline
$|\Psi^+\rangle$ & 0.5 & -0.5 & -0.5 & 0.5 \\ \hline
$|\Phi^-\rangle$ & 0.5 & 0.5 & -0.5 & -0.5 \\ \hline
$|\Psi^-\rangle$ & 0.5 & 0.5 & 0.5 & 0.5 \\ \hline
\end{tabular}
\caption{Optimal entanglement witnesses for the Bell states in the 
form $ W=C_I \mathbb{I}+C_x X_I\otimes X_S + C_y Y_I\otimes Y_S + C_z Z_I\otimes Z_S$.}
\end{table}
\end{center}

\section{Experiment}

NMR systems have been extensively used to test quantum information processing protocols.
Most of the experiments deal with entanglement, and therefore
the detection of entanglement by direct measurements is desirable. The main feature of NMR quantum 
information processing is the excelent control of unitary transformations over qubit states, provided 
by the use of radiofrequence pulses, which results in high fidelity.
Our experiment is performed on a liquid-state enriched carbon-13 Chloroform sample at room temperature
in a Varian 500MHz shielded spectrometer.
This sample exhibits two qubits encoded in the $^{1}H$ and $^{13}C$ 1/2-spin nuclei. The two qubit state 
is represented by a density matrix in the high temperature limit, which takes the form $\rho_{AB}$ $=$ $\mathbb{I}/4$ $+$ $\epsilon\Delta\rho_{AB}$,
where $\epsilon$ $=$ $\hbar\omega_L/4k_{B}T$ $\approx$ $10^{-5}$ is the ratio between the magnetic 
and thermal energies and $\Delta\rho_{AB}$ is the deviation matrix \cite{Oliveira}. Another form to write the 
density matrix of the NMR system is:
\begin{equation}\label{09a}
\rho_{AB} = \frac{1-\epsilon}{4}\mathbb{I} + \epsilon\rho_{1},
\end{equation}
\par\noindent
where $\rho_1$ is a density matrix. The matrix $\rho_1$ is directly related to the NMR observables, since 
$M^{\pm}$ $\propto$ \\
$Tr\left\{(I_{x}\pm iI_{y})\rho_{1}\right\}$.

By having the two entanglement witness well-defined, we can look at three classes of NMR states:
\begin{itemize}
\item entangled states which can be detected by $F$, 
\item entangled states which can not be detected by $F$,
\item separable states.
\end{itemize}

\begin{center}
\begin{figure}%
\includegraphics[scale=0.3]{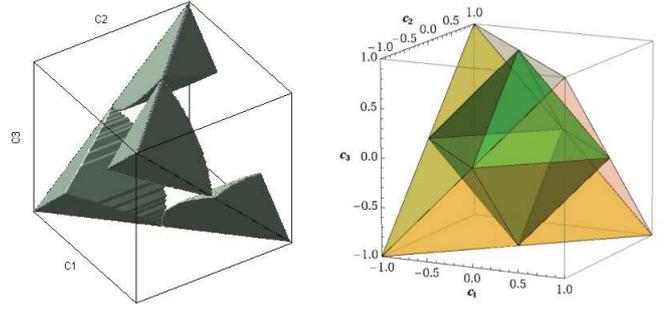}
\caption{In the righthand side is shown the geometry of the Bell-diagonal states, where the internal octahedron 
is the region of the separable states, with the entangled states being outside this tetrahedron. In the lefthand side, 
the Bell-diagonal states detected by F are in the shadowed volume. The empty volume represents the set of Bell-diagonal states 
such that $F$ $\geq$ $0$, i. e., the entangled states that are not detected by $F$ and the separable ones.  
The figure on the righthand side is credited to \cite{Laves}.}%
\label{detectionF}%
\end{figure}
\end{center}

These three classes of states will be considered using Bell-diagonal states, which are described 
by the equation:
\begin{equation}\label{09}
\rho = \frac{1}{4}\mathbb{I} + \sum_{i=1}^{3}c_{i}I_{i}\otimes I_{i},
\end{equation}
\par\noindent
where $c_{i}$ $\in$ $[-1, 1]$ and $I_i$ $=$ $2\sigma_i$, where $\sigma_i$ are the well known Pauli matrices. For this class of states,
the region of detection of entanglement by $F$ is given in the shadowed volume of Figure \ref{detectionF}. In the figure, the 
empty volume inside the lefthand side tetrahedron represents the $F$-non-detected states. The separable states being inside the octahedron in the 
righthand side figure. The entanglement witnesses $F$ and $W$ can detect the presence of entanglement not only for Bell-diagonal states, but for any class of states. In this paper, it will be shown the detection of entanglement by these two EW on the decoherence process of NMR, the relaxation.

The measurements of the magnetizations have been made by observing the NMR spectra directly. 
The NMR spectra of a two qubit spin-$1/2$ molecule gives the measurements of 
four projectors of the Hilbert-Schmidt space, by combining the intensities of the two lines of the spectra of each nucleus \cite{Mueller}.
The projectors measured in the first nucleus (with a very similar equation for the projectors measured in the second nucleus) 
are given by \cite{Mueller}:
\begin{equation}\label{09b}
\left(
\begin{array}{c}
Tr(\rho \tilde{I}_{-}\otimes\mathbb{I})\\
Tr(\rho \tilde{I}_{-}\otimes \tilde{Z})
\end{array}
\right) = 
\left(
\begin{array}{cc}
1 &  1\\
1 & -1
\end{array}
\right)
\left(
\begin{array}{c}
S(\omega_1 -\omega_{12})\\
S(\omega_1 +\omega_{12})
\end{array}
\right),
\end{equation}
\par\noindent where $\tilde{I}_{-}$ $=$ $U(X -iY)U^{\dagger}$ and $\tilde{Z}$ $=$ $UZU^{\dagger}$, with $U$ being a preparatory pulse that transforms
four desired observables of the Hilbert-Schmidt space basis into $I_{-}\otimes\mathbb{I}$ and $I_{-}\otimes Z$, the 
four basis elements observable in the NMR spectra. $S(\omega_1 -\omega_{12})$ and $S(\omega_1 +\omega_{12})$ are the line intensities.

The preparatory pulse that leads the desired basis elements into the observable ones is a $(\pi/2)$ pulse in the
$x$ or $y$ directions in one or both spins. For example, to read the two desired correlation functions, $\langle Z\otimes Z\rangle$
and $\langle X\otimes X\rangle$, the preparatory pulse necessary is a $(\pi/2)_{y}$ pulse in the second qubit, 
the $^{13}C$ nucleus in this case. These two correlation functions will be present in the second line of the lefthand side of Eq. (\ref{09b}), i. e.,
the difference between the normalized intensities in the NMR spectra of each nucleus. The correlation function 
$\left\langle X_{I}\otimes X_{S} \right\rangle$ is observed in the real part of the $^{1}H$ spectra, 
while $\left\langle Z_{I}\otimes Z_{S} \right\rangle$ is observed in the real part of the $^{13}C$ spectra.
The measurements were obtained after the phase adjustment, using the equilibrium state as reference, and the 
removal of the background signal present at the NMR spectra.

\begin{center}
\begin{figure}%
\includegraphics[scale=0.3]{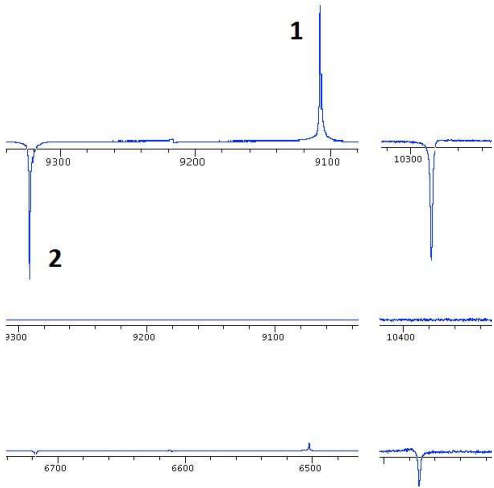}
\caption{The NMR spectra used to measure the correlation functions $\left\langle X_{I}\otimes X_{S} \right\rangle$ and
$\left\langle Z_{I}\otimes Z_{S} \right\rangle$ for each of the implemented states. The $^{1}H$ spectra is at lefthand side, 
while the $^{13}C$ spectra is at the righthand side. $(i)$ the NMR spectra for the $|\beta_{11}\rangle$ state. To 
illustrate the measurement of the correlation functions, $\left\langle X_{I}\otimes X_{S} \right\rangle$ is measured by looking
at the difference between the normalized line intensities of lines 1 and 2 at the lefthand side. $(ii)$ the NMR spectra for the identity. 
$(iii)$ the NMR spectra for the nondetected entangled state.}%
\label{nmrspectra}%
\end{figure}
\end{center}

\begin{center}
\begin{figure}%
\includegraphics[scale=0.45]{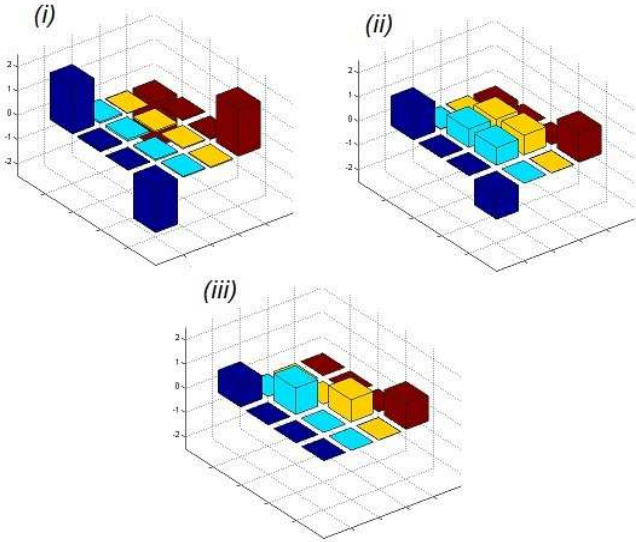}
\caption{Quantum state tomographies of the three states. The real parts are multiplied by a factor of $4$ 
to make the experimental errors more clear, and the imaginary parts are too small in all cases. $(i)$ the 
tomographed $|\Phi^-\rangle$ state. $(ii)$ the quantum state tomography of the identity, the separable 
state. $(iii)$ the tomographed Bell-diagonal state that is entangled and $F$-non-detected state.}%
\label{tomographies}%
\end{figure}
\end{center}

The $F$-detected entangled state which has been prepared is the $|\Phi^{-}\rangle$ $=$ $\frac{1}{\sqrt{2}}$
$\left(|00\rangle-|11\rangle\right)$ state, for which we measured $\left\langle Z_{I}\otimes Z_{S} \right\rangle$ $=$ $1.00\pm 0.01$ and 
$\left\langle X_{I}\otimes X_{S} \right\rangle$ $=$ $-1.01\pm 0.01$. 
These values give us $F$ $=$ $-0.51\pm 0.01$, in excelent agreement with the theoretical values, given by
$\left\langle Z_{I}\otimes Z_{S} \right\rangle$ $=$ $1.00$, $\left\langle X_{I}\otimes X_{S} \right\rangle$ $=$ $-1.00$ 
and $F$ $=$ $-0.50$. The Bell-diagonal state that is not detected 
by $F$ is given by $c_1$ $=$ $-0.20$, $c_2$ $=$ $1.00$ and $c_3$ $=$ $0.20$.  For this state, we measured 
$\left\langle Z_{I}\otimes Z_{S} \right\rangle$ $=$ $0.22\pm 0.01$ and $\left\langle X_{I}\otimes X_{S} \right\rangle$ 
$=$ $-0.17\pm 0.01$, which give us $F$ $=$ $0.15\pm 0.01$.  While the theoretical 
values are $\left\langle Z_{I}\otimes Z_{S} \right\rangle$ $=$ $0.20$ and $\left\langle X_{I}\otimes X_{S} \right\rangle$ $=$ $-0.20$, 
resulting in $F$ $=$ $0.14$. As an example of separable state, we prepared the identity, which is the state of maximum 
statistical mixture. For this state we measured $\left\langle Z_{I}\otimes Z_{S} \right\rangle$ $=$ $0.01\pm 0.01$ and 
$\left\langle X_{I}\otimes X_{S} \right\rangle$ $=$ $0.00\pm 0.01$, giving $F$ $=$ $0.25$ $\pm$ $0.01$. 
The theoretical expectation values of the correlation functions are null for the identity, with $F$ $=$ $0.250$. 
The fidelity of the tomographed states were found to be of order of $0.99$, $0.97$ and $0.98$, respectively. 
The tomographed states can be seen in the Figure \ref{tomographies}.

Since the entanglement witnesses $W$ need the measurement of one more projector, it is needed another reading pulse
to perform the measurement of $\langle Y_I\otimes Y_S\rangle$. In this case, the $\left(\pi/2\right)_x$ in 
the $^{13}C$ nucleus. The values of $\langle Y_I\otimes Y_S\rangle$ for the prepared states are given by
$0.96\pm 0.01$, $0.96\pm 0.01$ and $0.00\pm 0.01$, respectively. The theoretical values are, respectively, 
$1.00$, $1.00$ and $0.00$. For the $|\Phi^-\rangle$ state, it is needed to use the witness given by the 
values in the third line of the table \ref{tabela}, since the $F$-nondetected entangled state is at the same 
region of the tetrahedron (see Fig. \ref{detectionF}), the same entanglement witness is adequate to evaluate the 
entanglement for this state. The measured values for the entanglement witnesses are 
$W_{|\Phi^{-}\rangle}$ $=$ $-1.01$ $\pm$ $0.01$, for $|\Phi^{-}\rangle$,
$W_{|\Phi^{-}\rangle}$ $=$ $-0.20$ $\pm$ $0.01$, which shows that this state is detected by $W$. For the identity,
$W_{|\Phi^{-}\rangle}$ $=$ $0.50$ $\pm$ $0.01$.

The entanglement measured on each state is given by
$0.96$ $\pm$ $0.01$, $0.14$ $+$ $0.01$ and $0.00$ $\pm$ $0.01$, respectively. The quantification of entanglement 
have been made using the generalized robustness of entanglement \cite{Steiner, Cavalcanti, Brandao2}, a well-known measure of entanglement
based on the notion of ``distance'' between an entangled state and the set separable ones in the Hilbert-Schimdt space.
An advantage of this measure of entanglement is the fact that even for multipartite systems, it can identify
the various types of entaglement that these systems can exhibit.

\begin{center}
\begin{figure}%
\includegraphics[scale=0.25]{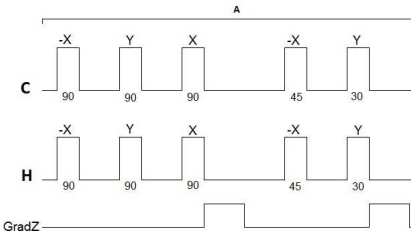}
\caption{Pulse sequence employed for the preparation of the entangled nondetected state. 
The first stage (A) is the passage from the thermal equilibrium state to the pseudo-pure 
state $|00\rangle$, after the GRAPE pulse (B), the pseudo-EPR (C), the reading pulse
(D) and the measurement. This is the sequence for the measurement in the $^{13}C$ nucleus. 
For measurements at the $^{1}H$ nucleus, the pulse sequence is the same, but with the 
corresponding lines of both nucleus interchanged. Above each box that indicates the 
radiofrequence pulses is the phase of the pulse, with the angle of rotation below the box.}%
\label{pulseseq}%
\end{figure}
\end{center}

The preparation of the $F$-detected and the identity states was made by using transfer gates \cite{Jones}, while the entangled and $F$-nondetected 
state was produced using the technique known as GRAPE \cite{Khaneja}, which takes the state $|00\rangle$ into the state 
$\sqrt{0.6}|00\rangle$ $+$ $\sqrt{0.4}|11\rangle$. A gradient pulse is applied 
to kill the coherences of this state, and the desired 
state is obtained after the application of a pseudo-EPR gate (see Figure \ref{pulseseq}). The 
quantum state tomography employed here was the variational quantum state tomography proposed by Maciel et al. \cite{Vianna}.

\begin{center}
\begin{figure}%
\includegraphics[scale=0.6]{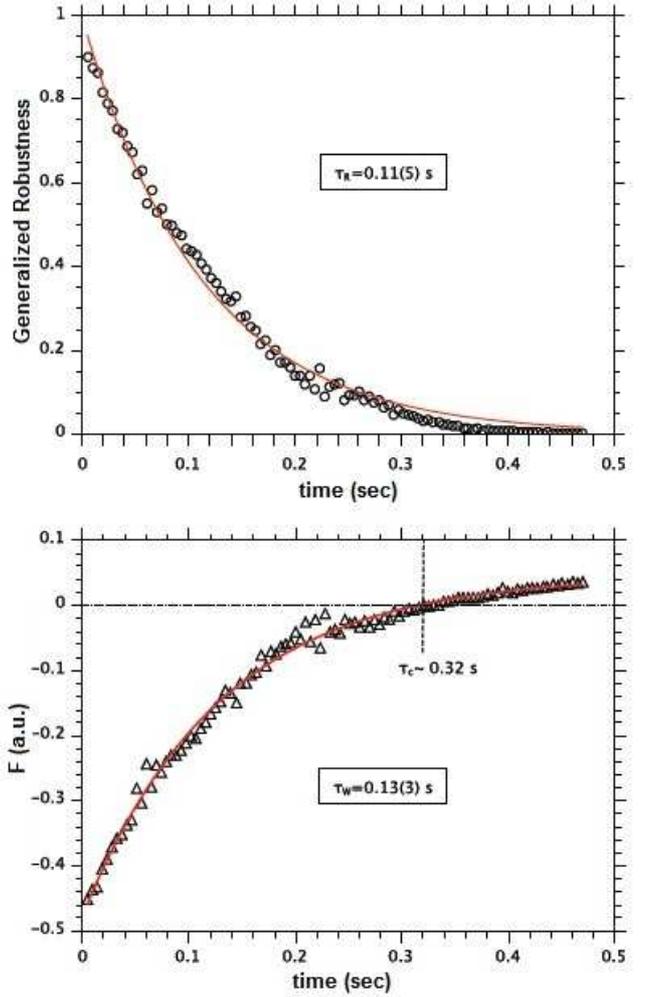}
\caption{Detection of entanglement by $F$ and generalized robustness under relaxation. The figure shows 
explicitly that there are states that are entangled but not detected by $F$ not only for Bell-diagonal states, since 
the transversal relaxation takes the state $|\Phi^-\rangle$ outside the class of the Bell-diagonal states.
This region is localized between the times near to $0.3s$ until approximately $0.4s$. At the figures, $\tau_R$ and $\tau_W$ 
are the characteristic times of the curves of detection of entanglement by the generalized robustness and by the EW, respectively.
The time $\tau_C$ is the approximate time which indicates the end of detection of the entanglement by the EW.
}%
\label{expcurves}%
\end{figure}
\end{center}

As an extension of the above study, the detection of entanglement by $F$ in the relaxation (decoherence) process, for the initial state
$|\Phi^{-}\rangle$, was studied. Using the generalized robustness \cite{Steiner} to quantify the entanglement, it is possible to
compare the detection of entanglement by the two methods.
As can be seen in the Figure \ref{expcurves}, the EW stops detecting 
the entanglement at the time $\tau_c$ $=$ $0.32$ seconds, near the transverse relaxation time of the hydrogen, given by $0.31(2)$ seconds for this sample,
while the generalized robustness still shows the presence of entanglement for a few miliseconds after this time.
But, as it is clear from the figure, the EW has a large range of detection during the decoherence. Another feature that comes from the 
data analysis of detection of entanglement by the generalized robustness and the detection by the EW
is that the entanglement decays with a characteristic time given by the the lowest decoherence time of the system, 
at this case, the transverse relaxation time of the $^{13}C$ nucleus, which is $0.11(2)$ seconds for this sample.

In Fig. 7, we compare the witnesses $F$ and $W$ for the entanglement
of the state $|\Phi^-\rangle$ under relaxation. In one hand, we see
that $W$ is more sensitive than $F$ and is a better {\em quantifier},
 but on the other hand they detect entanglement in the same region.
\begin{figure}
\includegraphics[scale=0.45]{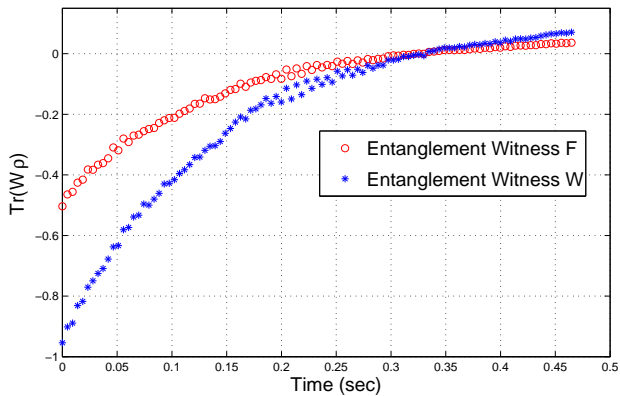}
\caption{Detection of entanglement by $F$ and $W$ under relaxation.
The two witnesses  detect entanglement in the same region,
but $W$ {\em quantifies}  entanglement better than $F$.} 
\end{figure}

\section{Conclusions}

In this paper, it has been made the experimental implementation of the EW proposed by Rahimi et al \cite{Rahimi} in the NMR context.
It was shown explicitly the detection of entanglement, by this EW, for two different situations. 
First, it has been shown the region of detection by the EW for the class of Bell-diagonal states, with examples of entangled states 
that are detected and are not detected by the EW and a separable state. It was also shown how is the detection of the entanglement
by the EW in the NMR decoherence proccess, the transversal relaxation. In this case, it was clearly shown the presence of a time interval
which has a little presence of entanglement that is not detected by the EW. Instead of the fact that 
the EW can not be used to quantify entanglement for the two classes of states studied in this paper, this EW
can detect entanglement for a large number of states in these two situations with the application of only one 
preparatory pulse to measure the EW. While the complete quantum state tomography,
necessary to calculate entanglement quantifiers such as the generalized robustness and the concurrence,
demands the application of four pulses to reconstruct the density matrix.

It was also proposed by a simple method other entanglement witnesses, $W$, that are optimal for each region of 
the entangled Bell-diagonal states. As could be seen by an example, these EW can detect the entangled Bell-diagonal states 
that are not detected by $F$. In the context of relaxation, the comparison of the two entanglement witnesses $W$ and $F$ 
shows that both detects the presence of entanglement in the same region, but $W$ with the advantage of a better quantification 
of entanglement in this region.

A possible next step would be the development of an EW that is optimall in the context of relaxation for a given state. 
Another advance would be the development in the context of NMR experiments of EW for the 
detection of multipartite entanglement for systems with larger number of qubits, where there is a considerable experimental cost 
to implement quantum state tomography.

\begin{acknowledgements}
J. G. Filgueiras thanks A. Gavini-Viana and A. C. Soares for helpfull discussions and 
for sharing some MATLAB codes. Financial support by Brazilian agencies
CAPES, CNPq, FAPERJ, FAPEMIG, and  INCT-IQ (National 
Institute of Science and Technology for Quantum Information).
\end{acknowledgements}

\end{document}